\documentstyle[aps,preprint,psfig]{revtex}
\begin{document}
\begin{center}
{\bf GLAUBER MODEL FOR HEAVY ION COLLISIONS FROM LOW ENERGIES TO 
  HIGH ENERGIES}
\footnote{Lectures notes for the physics batch of BARC Training School}
\end{center}
\begin{center}
P. Shukla \\
{\it Nuclear Physics Division \\
Bhabha Atomic Research Centre,Bombay 400 085 } \\
\end{center}

   The Glauber model is extensively applied to heavy
ion collision for describing a number of interaction processes over
a wide range of energies from near the Coulomb barrier to
higher energies.  The model gives the
nucleus-nucleus interaction in terms of interaction
between the constituent nucleons with a given density 
distribution.  The model is a semiclassical model picturing 
the nuclear collision in the impact parameter representation
where the nuclei move along the collision direction in a straight path.
 In these lectures we derive 
this model and discuss its applications in variety of 
problems in nuclear and high energy physics. 

\section{Cross sections and Impact parameter representation}

   The total reaction cross section in the collision of
two nuclei as per the partial wave analysis is given by 
\begin{equation}
\sigma_R=\frac {\pi}{k^2} \sum_{l=0}^{\infty} (2l+1) (1-|S_l|^2).
\end{equation}
Here $S_l=\exp(-2i\delta_l)$ is called the scattering matrix 
and $\delta_l$ is the nuclear phase shift.
The factor $(1-|S_l|^2)$ is called the transmission coefficient and
$|S_l|^2$ is referred to as transparency function or the 
probability that the projectile undergoes no interaction at
a given $l$.
 In a semiclassical approximation, one can write angular momentum
$l$ in terms of momentum $k$ and impact parameter $b$ as
\begin{equation}
 l+{1\over 2} =  k b.
\end{equation}
Thus, using $(2l+1)=2kb$ and$\sum_l = k \int db$ we can get the
reaction cross section in impact parameter representation as
\begin{equation}
\sigma_R= 2 \pi \int b db (1-|S(b)|^2).
\end{equation}
The factor $Tr(b)=1-|S(b)|^2$ is nothing but the transmission coeff. 
If the two nuclei are assumed to be sharp spheres 
with radii $R_1$ and $R_2$ then
\begin{eqnarray}
 Tr(b) &=& 1 \hspace{.1in}{\rm for} \hspace{.1in} b\le R_1+R_2 \nonumber\\
       &=& 0 \hspace{.1in}{\rm for} \hspace{.1in} b > R_1+R_2.
\end{eqnarray}
The total reaction cross section in this case is given by
\begin{equation}
\sigma_R= 2 \pi \int_0^{R_1+R_2} b db = \pi (R_1+R_2)^2.
\end{equation}
This is the well-known geometric formula for the cross section.

The nucleus-nucleus differential cross section as a function of 
center of mass scattering angle $\theta$ is described as
\begin{equation}
\frac{d\sigma}{d\Omega}= |f(\theta)|^2,
\end{equation}
where the scattering amplitude $f(\theta)$ for the non identical spinless
nuclei is given by partial wave analysis which can be written as a sum of 
Coulomb and nuclear scattering amplitudes as
\begin{eqnarray}
f(\theta) &=& f_C(\theta)+f_N(\theta) \nonumber \\
  &=& f_C(\theta)+\frac {1}{2ik} \sum (2l+1)(e^{2i\sigma_l})
         (S_l-1)P_l(\cos\theta).
\end{eqnarray}
Here $f_C(\theta)$, the Coulomb scattering amplitude
and  $\sigma_l$, the Coulomb phase shift are given by the following
expressions:
\begin{equation}
f_C(\theta)=- \frac{\eta}{2k} {\rm cosec}^2 \frac {\theta}{2} \,
     \exp\left[2i \sigma_0-2i \eta \ln \, \sin \frac{\theta}{2}\right],
\end{equation}
\begin{equation}
\sigma_{l+1}(\eta)=\sigma_{l}(\eta)+tan^{-1}(\frac{\eta}{l+1}).
\end{equation}
  The S wave Coulomb phase shift $\sigma_{0}$ can be set equal to zero
without any loss of generality.

\section{The Glauber model}

  The Glauber model  of multiple collision processes provides
a quantitative consideration of the geometrical configuration
of the nuclei when they collide. The Glauber model basically describe 
the nucleus-nucleus interaction in terms of elementary nucleon-nucleon 
interaction. It is based on the 
assumption that the nucleus travels in a straight line path.
At high energies this approximation is very good. At low 
energies the nucleus is deflected from straight line path 
due to Coulomb repulsion. The so called Coulomb modified
Glauber model (discussed in the next section) 
also gives very good description of heavy
ion scattering at low energies.

 Consider the collision of a projectile nucleus $A$ on a target
nucleus $B$.
 Define $t({\bf b}) d {\bf b}$ as the probability for having 
a nucleon-nucleon collision within the transverse area element
$d {\bf b}$ when one nucleon is situated at an impact parameter 
${\bf b}$ relative to another nucleon which is normalized according to
\begin{equation}
\int t({\bf b}) d {\bf b}=1.
\end{equation}
We define the probability of finding a nucleon in the volume
element $d{\bf b}_A dz_A$ in the nucleus $A$ at the position 
$({\bf b}_A, z_A)$ is 
$\rho_A({\bf b}_A, z_A) d{\bf b}_A dz_A$  which is normalized as
 \begin{equation}
\int \rho_A({\bf b}_A, z_A) d{\bf b}_A dz_A=1.
\end{equation}
Similarly, the probability of finding a nucleon in the volume
element $d{\bf b}_B dz_B$ in the nucleus $B$ at the position 
$({\bf b}_B, z_B)$ is 
$\rho_B({\bf b}_B, z_B) d{\bf b}_B dz_B$  which is normalized as
 \begin{equation}
\int \rho_B({\bf b}_B, z_B) d{\bf b}_B dz_B=1.
\end{equation}

\begin{figure}
\centerline{\psfig{figure=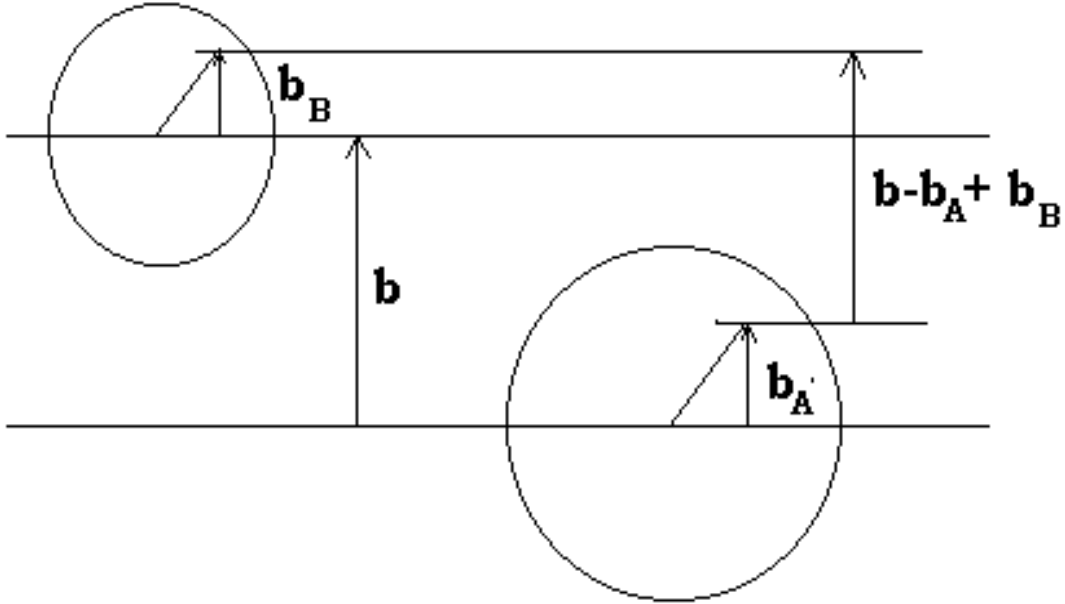,width=14cm,height=8cm}}
\caption{Collision of two nuclei at an impact parameter $b$}
\label{fig1}
\end{figure}

  The probability for occurrence of a nucleon-nucleon collision
[see Fig.~(\ref{fig1})]
when the nuclei $A$ and $B$ are situated at an impact parameter $\bf b$
relative to each other is given by
\begin{equation}
T(b)\sigma_{NN} = \int \rho_A({\bf b}_A, z_A) d{\bf b}_A dz_A \,\,
 \rho_B({\bf b}_B, z_B) d{\bf b}_B dz_B \,\,
 t({\bf b-b_A+b_B}) \,\, \sigma_{NN}.
\end{equation}
This can be written in terms of z-integrated densities as
\begin{equation}
T(b)\sigma_{NN} = \int \rho_A^z({\bf b}_A) d{\bf b}_A \,
 \rho_B^z({\bf b}_B) d{\bf b}_B \,
 t({\bf b-b_A+b_B}) \,\, \sigma_{NN}.
\end{equation}
Here $\sigma_{NN}$ is the inelastic nucleon nucleon cross section.
Thus, the collision probability we are talking about is for
an inelastic collision.
There can be upto $A\times B$ collision. The probability
of occurrence of $n$ collisions will be
\begin{equation}
 P(n,b) ={AB \choose n} (1-s)^n (s)^{AB-n} .
\end{equation}
Here, $s=1-T(b)\sigma_{NN}$.
The total probability for the occurrence of an inelastic event in
the collision of $A$ and $B$ at an impact parameter ${\bf b}$ is
\begin{eqnarray}
{d \sigma_{in}^{AB} \over db} = \sum_{n=1}^{AB} P(n,b)
 = 1 - s^{AB}.
\end{eqnarray}
  The total inelastic cross section can be written as
\begin{equation}
\sigma_{in}^{AB}=2 \pi \int b db \left( 1-s^{AB} \right).
\end{equation}
From here one can read the scattering matrix as
\begin{eqnarray}
|S(b)|^2 = s^{AB}=(1-T(b)\sigma_{NN})^{AB}.
\end{eqnarray}
In the optical limit, where a nucleon of projectile undergoes 
only one collision in the target nucleus 
\begin{eqnarray}\label{sb}
|S(b)|^2 \simeq \exp(-T(b)\sigma_{NN} AB).
\end{eqnarray}
The scattering matrix can be defined in terms of eikonal
phase shift $\chi(b)$ as
\begin{eqnarray}\label{sb2}
S(b)= \exp\left(-i\chi (b)\right).
\end{eqnarray}
Comparing Eq.~(\ref{sb}) with Eq.~(\ref{sb2}),
the imaginary part of eikonal phase shift is given by
\begin{eqnarray}
{\rm Im} \chi (b) 
  =T(b)\sigma_{NN} AB/2.
\end{eqnarray}
If the ratio of real to imaginary part of NN scattering 
amplitude is $\alpha_{NN}$ then real part of $\chi(b)$ is
\begin{eqnarray}
{\rm Re} \chi (b) 
  =T(b) \alpha_{NN} \sigma_{NN} AB/2.
\end{eqnarray}
Once we know the phase shift and thus the scattering matrix, we can
calculate the reaction cross section and the angular distribution.

\section{Calculation of $T(b)$ in momentum space}

In the co-ordinate space $T(b)$ is derived as
\begin{equation}
T(b) = \int \rho_A^z({\bf b}_A) d{\bf b}_A \,\,
 \rho_B^z({\bf b}_B) d{\bf b}_B \,\,
 t({\bf b-b_A+b_B}).
\end{equation}
It is a four dimensional integration: two over ${\bf b_A}$
and two over ${\bf b_B}$. It is convenient to write it
in momentum space as 
\begin{equation}
T(b) = {1\over (2\pi)^2}
  \int \rho_A^z({\bf b}_A) d{\bf b}_A \,\,
  \rho_B^z({\bf b}_B) d{\bf b}_B  \,\,
  \exp\left(-i{\bf q}.({\bf b}-{\bf b_A}+ {\bf b_B})\right) \,\,
   f_{NN}(q) d^2 q.
\end{equation}
Here $f_{NN}(q)$ is the $q$ dependence of NN scattering amplitude
given by
\begin{equation}
t({\bf b}) = {1\over (2\pi)^2}
  \int e^{-i{\bf q.b}} f_{NN}(q) d^2 q
\end{equation}
\begin{eqnarray}
T(b)&=& {1\over (2\pi)^2}  \int \exp(-i{\bf q.b}) \,\,
    \rho_A^z({\bf b_A}) \exp(i{\bf q.b_A}) d{\bf b}_A  \,\,
      \rho_B^z({\bf b_B}) \exp(-i{\bf q.b_B}) d{\bf b}_B \,\,
      f_{NN}(q) d^2 q   \nonumber \\
    &=& {1\over (2\pi)^2} \int e^{-i{\bf q.b}} 
       S_{A}({\bf q}) S_{B}(-{\bf q})  f_{NN}(q) d^2 q \nonumber \\
    &=& {1\over 2\pi}
       \int J_0(qb) S_{A}({\bf q}) S_{B}(-{\bf q})  f_{NN}(q) q dq.
\end{eqnarray}
Here $S_{A}(q)$ and  $S_{B}(-q)$ are 
the fourier transforms of the nuclear densities.
The function $f_{NN}(q)$ is the fourier transform of the
profile function $t({\bf b})$. The profile function for the NN
scattering can be taken as delta function if the nucleons
are point particles. In general it is taken as a gaussian function
of width $r_0$ as
\begin{equation}
t({\bf b}) = {\exp(-b^2/r_0^2) \over \pi r_0^2}.
\end{equation}
Thus, 
\begin{eqnarray}
f_{NN}(q) &=& \int e^{i{\bf q.b}} t({\bf b}) d{\bf b} \nonumber\\
&=&{1\over \pi r_0^2 } \int e^{i{\bf q.b}} \exp(-b^2/r_0^2) d{\bf b} \nonumber\\
&=& \exp(-r_0^2 q^2/4).
\end{eqnarray}
 Here, $r_0^2=0.439$ fm$^2$ is the range parameter and
$\sigma_{NN}$ is the nucleon-nucleon inelastic cross section which
is taken as 3.2 fm$^2$ at high energies.

\subsection{Calculation of $T(b)$ using Gaussian densities}
 If the nuclear densities are assumed to be of the Gaussian shape given by
\begin{equation}
\hspace{1in} \rho_i(r)=\rho_i(0) \exp(-r^2/a_i^2) \hspace{.5in}  (i=1,2).
\end{equation}
Here the parameters $\rho_i(0)$ and $a_i$ are adjusted to reproduce 
the experimentally determined surface texture of the nucleus.
The $z$-integrated density will be
\begin{equation}
\rho^z_i (r)=\rho_i(0)\sqrt{\pi} a_i \exp(-b^2/a_i^2).
\end{equation}
 This one can write in the momentum representation as
\begin{equation}
S_i (q)=\rho_i(0)(\sqrt{\pi} a_i)^3 \exp(-q^2 a_i^2/4).
\end{equation}
The overlap integral $T(b)$ can be written as 
\begin{eqnarray}
T(b) &=& {1\over (2\pi)^2} \int e^{-i{\bf q.b}} 
       S_{1}({\bf q}) S_{2}(-{\bf q})  f_{NN}(q) d^2 q \nonumber \\
     &=& {1\over (2\pi)^2} \pi^3 \rho_1(0) \rho_2(0) a_1^3 a_2^3
         \int e^{-i{\bf q.b}} \exp(-a^2 q^2/4) d^2 q,
\end{eqnarray}
where $a^2=a_1^2+a_2^2+r_0^2$. Performing $q$ integration we get
\begin{eqnarray}
T(b) &=& {1\over (2\pi)^2} \pi^3 \rho_1(0) \rho_2(0) a_1^3 a_2^3
        \left[{ 4\pi\over a^2} \exp(-b^2/a^2)\right] \nonumber \\
     &=& \pi^2 {\rho_1(0) \rho_2(0) a_1^3 a_2^3 \over a_1^2+a_2^2+r_0^2}
       \exp(-b^2/(a_1^2+a_2^2+r_0^2)).
\end{eqnarray}

\subsection{Calculation of $T(b)$ using 2pf densities}
  The two parameter fermi density is given by
\begin{equation}
\rho(r)=\frac{\rho_0}{1+\exp(\frac{r-c}{d})},
\end{equation}
  where $\rho_0=3/\left(4\pi c^3(1+\frac{\pi^2d^2}{c^2})\right)$.
Thus, the momentum density can be derived as
\begin{equation}
S(q)=\frac{8\pi\rho_0}{q^3} \frac{z e^{-z}}{1-e^{-2z}}
     \left( \sin x  \frac{z(1+e^{-2z})}{1-e^{-2z}}-x \cos x \right).
\end{equation}
  Where $z=\pi dq$ and $x=cq$. Here $d$ is the diffuseness and
$c$, the half value radius in terms of rms radius $R$ for
the 2pf distribution is calculated by $c=(5/3R^2-7/3 \pi^2 d^2)^{1/2}$.
 The equation  can be solved numerically for this density and the
overlap integral can be extracted.

\subsection{NN scattering parameters $\sigma_{NN}$ and
                 $\alpha_{NN}$ }
 In this section, we give the expressions for $\sigma_{NN}$ 
and $\alpha_{NN}$ obtained from the parameterization of 
experimentally measured cross section. The average $\sigma_{NN}$ 
in terms of proton numbers ($Z_P$ and $Z_T$) and neutron number 
($N_P$ and $N_T$) of projectile and target nuclei 
is written as
\begin{equation}
\sigma_{NN}={N_P N_T \sigma_{nn} + Z_P Z_T \sigma_{pp} 
    + (Z_P N_T + N_P Z_T) \sigma_{np}  \over A_P A_T}.
\end{equation}
Here, pp cross section $\sigma_{pp}$ and  
nn cross section $\sigma_{nn}$ are given in (fm$^2$) by 
\begin{equation}
\sigma_{pp}=\sigma_{nn}=1.373-1.504/\beta+0.876/\beta^2+6.867 \beta^2.
\end{equation}
Here $\beta$ is the velocity of projectile nucleon.
 For np cross section $\sigma_{np}$, two expression are used.
If the energy per nucleon $E_N > 10$ MeV, then 
\begin{equation}
 \sigma_{np}=-7.067-1.818/\beta+2.526/\beta^2+11.385 \beta.
\end{equation}
For $E_N < 10$ MeV, 
\begin{equation}
 \sigma_{np}={273 \over (1.-0.0553 E_N)^2 +0.35 E_N } 
        + {1763 \over (1.+0.334 E_N)^2+6.8 E_N}.
\end{equation}
The average $\alpha_{NN}$ is written as
\begin{equation}
 \alpha_{NN}={N_P N_T \alpha_{nn} \sigma_{nn} 
     + Z_P Z_T \alpha_{pp} \sigma_{pp}
     +(Z_P N_T + N_P Z_T) \alpha_{np} \sigma_{np} 
     \over  \sigma_{NN} A_P A_T }.
\end{equation}
The parameterized forms of $\alpha_{pp}$, $\alpha_{nn}$ and
$\alpha_{np}$ are written as
\begin{equation}
\alpha_{pp}= \alpha_{nn} = 0.1810+4.0818 p+0.3327 p^2
\end{equation}
and 
\begin{equation}
 \alpha_{np}=-.698+4.9762 p-1.277 p^2,
\end{equation}
where $p$ is the momentum of projectile nucleon in GeV.

\section {COULOMB MODIFIED GLAUBER MODEL}
  In the nucleus-nucleus collision at low energies 
it is the surface region of the nucleus that contributes 
to the scattering amplitude in a
non-trivial manner. Since $V/E<<1$ even for low and intermediate 
energies in the surface region, the eikonal approximation can be
extended to the low energy heavy ion collisions. 
  The basic assumption of the Glauber model is the description
of the relative motion of the two nuclei in terms of straight line
trajectory. The density overlaps are evaluated along straight lines 
associated with each impact parameter $b$. The modification in the 
straight line trajectory due to the Coulomb field can not be ignored,
especially in the case of heavily charged systems at relatively
low bombarding energies. 

\newpage
\begin{figure}
\centerline{\psfig{figure=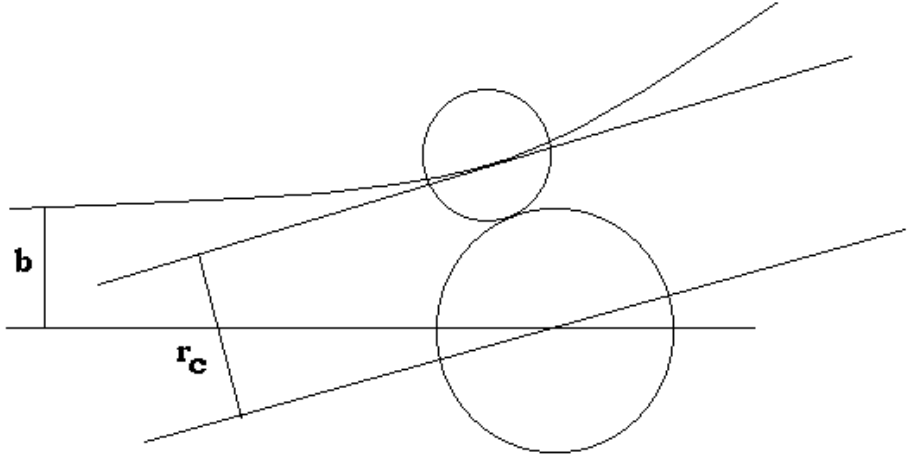,width=12cm,height=6cm}}
\caption{The straight line trajectory is assumed at the distance 
of closest approach $r_{c}$}
\label{fig2}
\end{figure}

  For low energy  heavy ion 
reactions the straight line trajectory is assumed at the distance 
of closest approach $r_{c}$ [see Fig.~(\ref{fig2})]
calculated under the influence of the 
Coulomb potentials for each impact parameter $b$ as given by,
\begin{equation}
 r_c=(\eta +\sqrt{\eta^2+ b^2 k^2})/k,
\end{equation}
which is a solution of the equation 
\begin{equation}
E-\frac{Z_1Z_2 e^2}{r}-\frac{\hbar^2 k^2}{2\mu}\frac{b^2}{r^2} = 0,
\end{equation}
\begin{equation}
E-\frac{Z_1 Z_2 e^2}{r}-E \frac{b^2}{r^2} = 0,
\end{equation}
\begin{equation}
k r^2-2 \eta r - b^2 k = 0.
\end{equation}
Here $\eta$ is the dimensionless Sommerfeld parameter defined as
\begin{equation}
 \eta=\frac {Z_1 Z_2 e^2}{\hbar v} = \frac {Z_1 Z_2 e^2}{E}{k\over 2}.
\end{equation}

Thus finally the reaction cross section will be
\begin{equation}
\sigma_{in}^{AB}=2 \pi \int b db \left( 1-(1-T(r_c)\sigma_{NN})^{AB} \right).
\end{equation}

\begin{figure}
\centerline{\psfig{figure=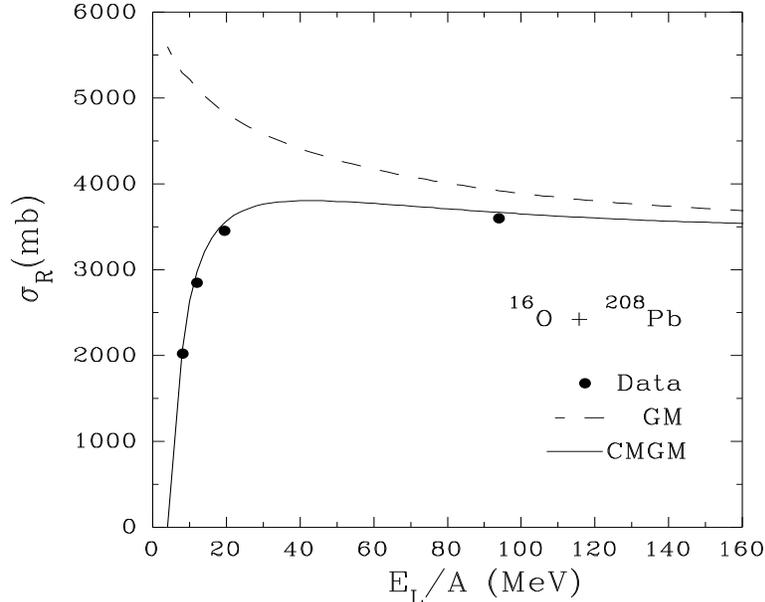,width=10cm,height=8cm}}
\caption{Reaction cross section for $^{16}$ O + $^{208}$ Pb system
     calculated with Glauber model (GM) and Coulomb modified 
   Glauber model (CMGM) along with the data}
\label{opb}
\end{figure}

Figure~(\ref{opb}) shows the 
reaction cross section for $^{16}$ O + $^{208}$ Pb system
as a function of projectile energy per nucleon 
calculated with Glauber model (GM) and Coulomb modified 
Glauber model (CMGM) along with the data. At higher energies
both GM and CMGM result merge. But at lower energies one
requires to take Coulomb modification in the trajectory.

\section{Participant-spectator picture in high energy heavy ion collisions}

\begin{figure}
\centerline{\psfig{figure=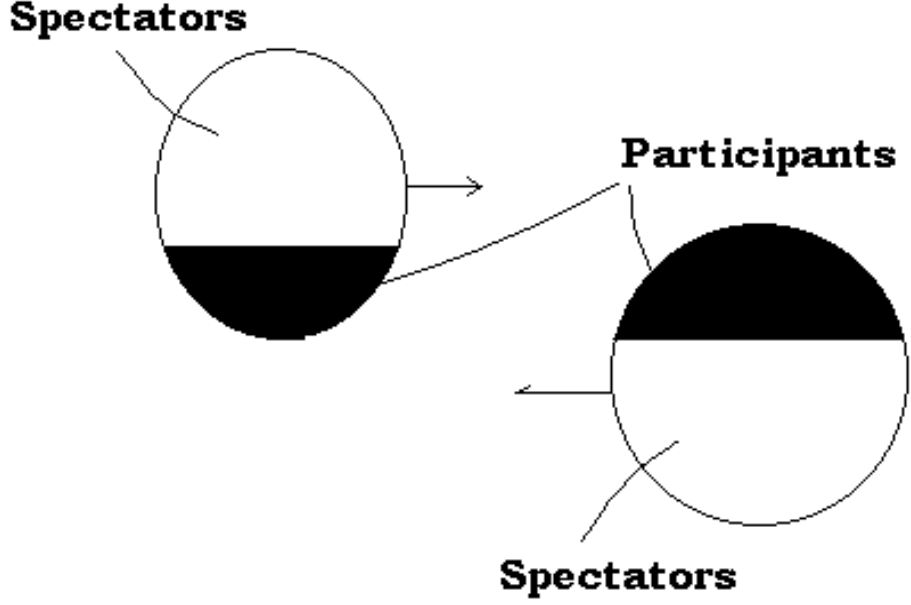,width=12cm,height=8cm}}
\caption{Spectator participant picture of heavy ion collision}
\label{fig3}
\end{figure}

 Nucleus-nucleus collisions at ultrarelativistic energies are being
explored in order to search for the formation of quark gluon plasma.
Already several experimental results are available. One of the stategies
is to describe these results as the superposition of the nucleon-nucleon
collisions and look for possible departures from this prescription. 
When two nuclei collide at these energies, the nucleons 
which come in the overlap region depending on the impact 
parameter are called participant nucleons and those which
do not participate are called spectators. [see Fig.~(\ref{fig3})]

\newpage

 These spectators and participants nucleon decide how much energy
is going in the forward and how much in the transverse
direction which can be measured. From these measured energies
one can know the impact parameter for a particular event.
Let us write the total inelastic cross section for a 
nucleus-nucleus collision as
\begin{equation}
\sigma_{in}^{AB}=2 \pi \int b db \left( 1-(1-T(b)\sigma_{NN})^{AB} \right).
\end{equation}
The term $(1-T(b)\sigma_{NN})^{AB}$ gives the probability that
in a nucleus nucleus collision none of the nucleons collided with
each other.
For nucleon-nucleon collision $A=B=1$ thus 
$s=(1-T(b)\sigma_{NN})$ gives the probability that two nucleon
at an impact parameter $b$ do not collide.

When two nuclei $A$ and $B$ collide the probability of nucleon remaining in the 
projectile will be
\begin{equation}
P_P=s(b)^B
\end{equation}
 and the probability of nucleon remaining in the target will be
\begin{equation}
P_T=s(b)^A.
\end{equation}

  The probability of
having $\alpha$  participant nucleons from the nucleus $A$ is given by
binomial distribution as  
\begin{equation}
 P(\alpha,b) ={A \choose \alpha } (1-s^B)^{\alpha} (s^B)^{A-\alpha}.
\end{equation}  
Similarly,
\begin{equation}
 P(\beta,b) ={B \choose \beta } (1-s^A)^{\beta} (s^A)^{A-\beta} .
\end{equation}
 The number of collisions will be
\begin{equation}
 P(n,b) ={AB \choose n} (1-s)^n (s)^{AB-n} .
\end{equation}
  The average number of projectile participant and its
standard deviation for each impact parameter $b$ is given by
\begin{equation}
<\alpha>=A[1-s(b)^B],
\end{equation}
\begin{equation}
\sigma_{\alpha}^2=A[1-s(b)^B] s(b)^B .
\end{equation}
  The average number of target participant and its
standard deviation for each impact parameter $b$ is given by
\begin{equation}
<\beta>=B[1-s(b)^A],
\end{equation}
\begin{equation}
\sigma_{\beta}^2=B[1-s(b)^A] s(b)^A.
\end{equation}
  The average number of total participant 
from both projectile and target for each impact
parameter is given by
\begin{equation}
N_{\rm participants}=A[1-s(b)^B]+B[1-s(b)^A].
\end{equation}
  The average number of N-N collisions is given by
\begin{equation}
N_{collisions}=AB[1-s(b)].
\end{equation}

\begin{figure}
\centerline{\hbox{
\psfig{figure=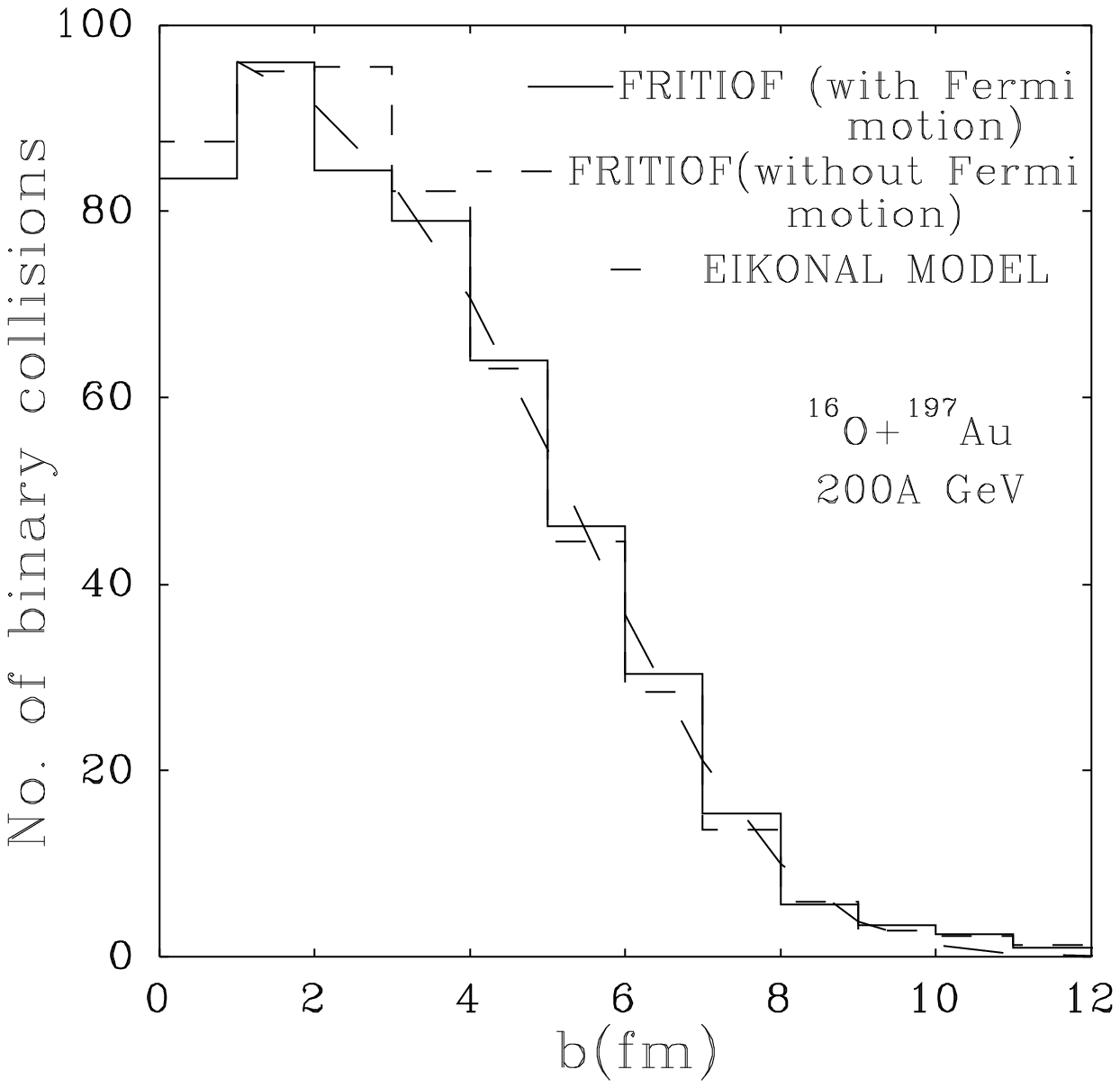,height=7.0cm,width=7.0cm} 
\psfig{figure=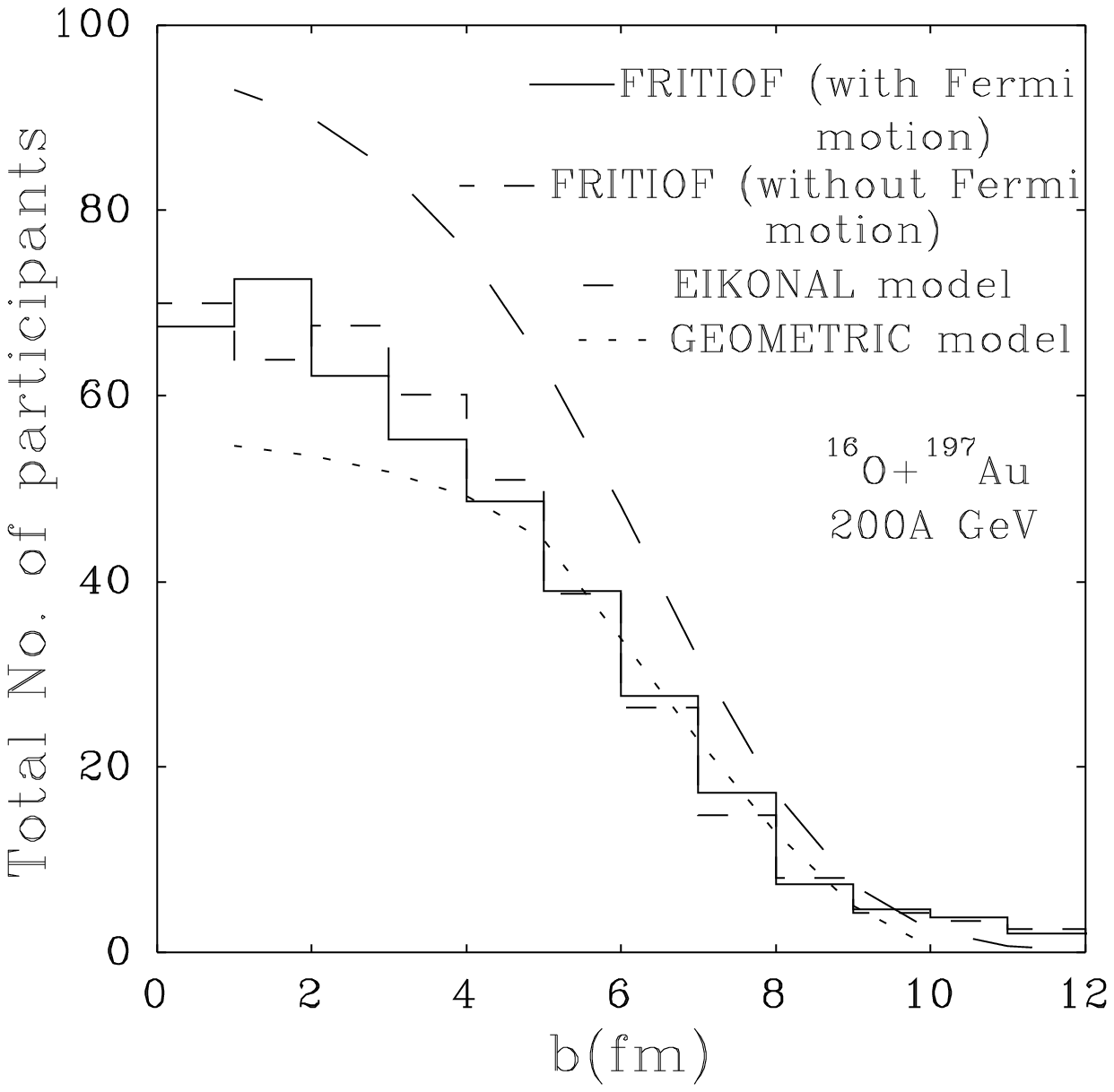,height=7.0cm,width=7.0cm} 
}}
\caption{\sf (a) The number of binary collisions as a function of
 impact parameter calculated with the
 expressions given in the text along with the Monte Carlo code 
 FRITIOF and Geometric model of J. Gosset et al.  
 (b) The total number of participants as function of
  impact parameter.}
\label{coll}
\end{figure}

The Fig.~(\ref{coll}) shows the number of participants and the number of 
N-N collisions as a function of impact parameter with the
expressions given in the text along with the calculations with
the Monte Carlo code FRITIOF and Geometric model of J. Gosset et al.  

\subsection{Forward energy}
The forward energy is measured in the detector called zero 
degree calorimeter (ZDC). The energy measured in the ZDC is
can be related to the number of spectators as
\begin{equation}
E_Z(\alpha)=(A-\alpha)E_0,
\end{equation}
where $E_0$ is the projectile energy per nucleon in the lab frame.

  The cross section of $\alpha$ nucleons participating from projectile is 
\begin{equation}
\sigma_{\alpha}=2 \pi \int P(\alpha,b) b db.
\end{equation}
 The forward energy flow cross section then can be written as
\begin{equation}
\frac{d\sigma}{dE_Z}=\frac{\sigma_{\alpha}}{E_0},
\end{equation} 
 taking into account the discrete nature of the variable $\alpha$.
The average forward energy for a collision is written as
\begin{equation}
E_Z(\alpha)=E_0(A-<\alpha>)=E_0 A[1-(1-s^B)]=E_0 A s^B.
\end{equation}

\subsection{Excitation energy}
  When the nucleons collide at relativistic energies they become excited
nucleons which due to time dilation become long-lived and decay into
secondary fragments outside the nuclei. This energy which
goes in exciting the nucleons is called excitation energy which manifests 
in the measurement of transverse energy and multiplicity. 
The (maximum) excitation energy which is related to the
number of participants is written as
\begin{equation}
E_{ex}=E_{cm}-m_N (\alpha+\beta).
\end{equation}
  The centre of mass energy for this case is given as
\begin{equation}
  E_{cm}^2 = ( (\alpha E_A + \beta E_B)^2- (\alpha {\bf P}_A + 
\beta {\bf P}_B)^2),            
\end{equation}
where
\begin{equation}
P_A=\sqrt{E_A^2-m_N^2} \hspace{.5in} {\rm and} \hspace{.5in}
     P_B=\sqrt{E_B^2-m_N^2}.
\end{equation}

 The cross section for having $\alpha$ participants from $A$ and $\beta$
participants from $B$ is 
\begin{equation}
 \sigma_{\alpha \beta} =2 \pi \int P(A,\alpha,b) P(B, \beta,b) b db. 
\end{equation}
 Here $E_A$ and ${\bf P}_A$ are the total energy and momentum of the 
nucleus $A$ per nucleon and $m_N$ is nucleon mass. 
The cross sections for all the possible combinations of
$\alpha$ and $\beta$ are calculated and  the results are binned to obtain the
cross section for each of the excitation energy bin. 

\subsection{Multiplicity}

  The total charged particle multiplicity in pp collision for a
center of mass energy $\sqrt s$ is given by 
\begin{equation}
<n_{ch}>=0.88+0.44 \ln(s)+0.118 (\ln(s))^2.
\end{equation}
  The center of mass energy for pair of participants from nucleus
$A$ and nucleus $B$ can be written as
\begin{equation}
s_{\alpha \beta}=\left( \frac{2 E_{cm}}{\alpha+\beta} \right) ^2.
\end{equation}
  Thus, the total multiplicity in nucleus-nucleus collision in
the case of $\alpha$ nucleons participating from nucleus
$A$ and $\beta$ participating from nucleus $B$ can be written as
\begin{equation}
<M>=\left( 0.88+0.44 \ln(s_{\alpha \beta})+0.118 (\ln(s_{\alpha \beta}))^2 
\right)\left(\frac{\alpha+\beta}{2}\right).
\end{equation}
The cross sections and the
multiplicity for all the possible combinations of projectile
and  target participants  are calculated and the results are
binned to obtain the cross section for each of the multiplicity bin.

\section{$j/\psi$ production cross section}

The probability for $j/\psi$ production in a nucleus nucleus collision
can be found out if we know the 
$j/\psi$ production cross section in a nucleon nucleon collision,
\begin{equation}
P_{j/\psi}^{AB}(b)= A B \sigma_{j/\psi}^{NN} T(b),
\end{equation}
The cross section is
\begin{equation}
\sigma_{j/\psi}^{AB}=A B \sigma_{j/\psi}^{NN}
 \int d{\bf b} T(b).
\end{equation}
The integral is 1 as per the normalization condition
\begin{equation}
{\sigma_{j/\psi}^{AB}\over \sigma_{j/\psi}^{NN}} =AB.
\end{equation}
The $j/\psi$ once produced can collide with the other 
nucleons present in the nucleus and thus 
it is possible that it is absorbed (converted 
into $D(c\bar u)$ and $\bar D (\bar c u)$)
during the path it travels in the nucleus. 
This nuclear absorption can be taken into account by writing
\begin{equation}
{\sigma_{j/\psi}^{AB}\over \sigma_{j/\psi}^{NN}} =
 AB \exp(-L\sigma_{abs} \rho_0).
\end{equation}
Here, $\rho_0=0.14$ fm$^{-3}$ is the equilibrium nuclear matter density,
$\sigma_{abs}=6.2$ mb is the absorption cross section and $L$ is the
average path length given by
\begin{equation}
L=L_A + L_B = {3\over 4} \left( {A-1\over A} R_A + {B-1\over B} R_B\right),
\end{equation}
where, $R_A=r_0 A^{1/3}$ and $R_B=r_0 B^{1/3}$.

\begin{figure}
\centerline{\psfig{figure=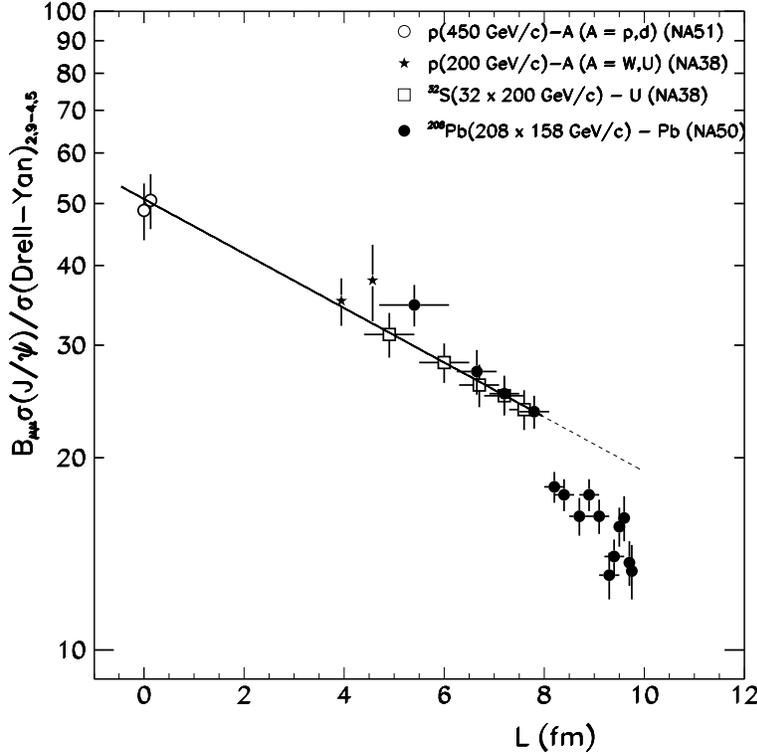,width=10cm,height=10cm}}
\caption{The suppression of $J/\Psi$ over Drell-Yan pairs in
$2.9\,{\rm GeV}< M <4.5$ GeV as a function of the average nuclear path length 
$L$ of the  $c\bar c$ pre-resonance, for NA38, NA50 and 
NA51 data. The NA50 Pb+Pb data (filled circles) is from the 1996 run.}
\label{jpsi}
\end{figure}

 Figure~(\ref{jpsi}) shows the suppression of $J/\Psi$ over Drell-Yan pairs 
as a function of the average nuclear path 
length $L$ of the  $c\bar c$ pre-resonance, for NA38, NA50 and 
NA51 data. 

\section{The Monte Carlo event generators}
There are many Monte Carlo modes available to study the
relativistic heavy ion collisions as superposition of nucleon-nucleon
collisions such as FRITIOF and IRIS. 
  In these codes the two nuclei are made to interact at a random
impact parameter. A Woods-Saxon distributed nuclear density,
giving the probability of finding a nucleon at a given distance
from the centre of the nucleus, is used. Each nucleon is given
space coordinates and the number of subcollisions is calculated 
letting a projectile hit the nucleus at a random impact parameter.
Here a frozen straight line geometry is used and all nucleons
inside a cylinder surrounding the path of the projectile are
considered to participate. The number of
binary collisions is recorded and the two partners involved
in each subcollision are determined. In each subcollision,
momentum is exchanged.

 FRITIOF is based on the LUND string picture of hadron-hadron
interactions. In this picture, each nucleon-nucleon collision
results in excitation of the nucleon by the stretching of a 
string between the valence quark and diquark.
In nucleus-nucleus interactions each nucleon can make several encounters,
the objects can get further excited, thereby increasing their masses
during the passage through the nucleus.
 A phenomenological excitation function determines the mass and momentum 
of the string after each interaction. 
 Finally, all the excited objects hadronize independently, like 
massless relativistic strings, according to the Lund model for the jet 
fragmentation. The hadronization takes place outside the nuclei
and thus no intranuclear cascading is considered.

  Three other models based on string picture of hadron-hadron
interactions are IRIS, MCFM and VENUS all of which are colour
exchange models based on the Dual Parton Model (DPM) of Capella
et al. Here the basic mechanism of string formation is colour
exchange between the quarks of the colliding nucleons. In these 
models the string properties can be calculated from structure
functions.  HIJET is an extension of the ISAJET model of 
hadron interactions and MARCO is based upon a phenomenological
parameterization of nucleon-nucleon collisions. ATTILA has the
possibility of rope formation from overlapping strings and MCFM
and HIJET which allow cascading of produced particles when the
assumed particle formation time is short. RQMD is an extension
to relativistic energies of the Quantum Molecular Dynamics which
has been applied to nucleus-nucleus collisions at much lower
energies.
   
   The nucleus-nucleus collision geometry should be the same
for all of the Monte-Carlo codes since nuclear density 
distributions are well-known from nuclear physics and should
not be treated as free parameters. It should be noted that 
deviations as small as 5-10 \% in the treatment of the nuclear
geometry are significant since one hopes to draw conclusions 
about deviations from the measurements which are of similar
magnitude. It is clear that before firm conclusions can be drawn 
upon the significance of the differences in the physics of the 
models at the nucleon-nucleon level, or on whether there exists
experimental evidence for rescattering of the produced
particles, or even for QGP formation, it will be necessary to
ensure that the various models treat the nuclear geometry
correctly and consistently.

\end{document}